\documentclass[twocolumn, prl, showpacs]{revtex4}

\usepackage{epsfig}

\usepackage{amsmath}
\usepackage{amsfonts}
\usepackage{amssymb}

\begin{document}

\title{Extended navigability of small world networks: exact results and new insights}

\author{C\'ecile Caretta Cartozo} \author{Paolo De Los Rios}
\affiliation{Institute of Theoretical Physics, SB, Ecole
Polytechnique F\'ed\'erale de Lausanne (EPFL), CH-1015, Lausanne,
Switzerland}

\date{\today} 

\begin{abstract}

Navigability of networks, that is the ability to find any given
destination vertex starting from any other vertex, is crucial to their
usefulness. In 2000 Kleinberg showed that optimal navigability could
be achieved in {\it small-world} networks provided that a special
recipe was used to establish long range connections, and that a greedy
algorithm, that ensures that the destination will be reached, is
used. Here we provide an exact solution for the asymptotic behavior of
such a greedy algorithm as a function of the system's parameters. Our
solution enables us to show that the original claim that only a very
special construction is optimal can be relaxed depending on further
criteria, such as, for example, cost minimization, that must be
satisfied.

\end{abstract}


\maketitle

By endowing nodes with both well-connected local neighborhoods
  and long-range shortcuts, that dramatically reduce the distances to
  any other node, transport on small-world (SW) networks is both
  locally and globally efficient~\cite{Watts1998}. Such feature has
  made SW networks appealing for several fields, such as social,
  computer and life
  sciences~\cite{Watts2002,Dodds2003,Liben-Nowell2005,Menczer2002,Lomholt2005,Sporns2004,Bassett2006}.
  Unfortunately, taking full advantage of the small node-to-node
  distances requires a global knowledge of the system, that is, in general,
  not accessible.  It is thus important to devise decentralized
  algorithms that rely only on local information and that are able to
  find good, although sub-optimal, routes from source to
  destination~\cite{Kleinberg2006}. The analysis of a prototype
  decentralized algorithm showed that the precise recipe used to
  establish the long-range shortcuts affects the ability of
  decentralized algorithms to navigate the
  networks~\cite{Kleinberg2000}.  Here we provide an exact solution in
  any dimension for such problem, and, in light of recent insights
  into the properties of SW networks~\cite{Petermann2006} and in
  contrast to the findings in~\cite{Kleinberg2000}, we show that there
  is a broad range of SW networks that are optimally navigable.
  
SW networks can be obtained from regular lattices in $d$ dimensions by
adding to every node, with probability $q$, a long-range connection to
another node taken at random over the whole
lattice~\cite{Newman1999a}. The key feature of SW networks is that the
average distance between any two nodes grows at most logarithmically
with the linear size $L$ of the lattice~\cite{Newman1999a}.  In order
to analyze how different shortcut addition schemes affect the
performances of a decentralized algorithm,
Kleinberg~\cite{Kleinberg2000} specified an additional rule for the
selection of the long-range partnerships: the probability $s(l)$ that
the shortcut added to a node ends at a node at euclidean (or lattice)
distance $l$ is a decaying power-law, $s(l) = \mathcal{N}
l^{-\alpha}$, where $\mathcal{N}$ is the normalization over the whole
lattice.  The two key ingredients for the construction of the SW
network are thus the shortcut addition probability $q$ and the
exponent $\alpha$.

\begin{figure}[h!] 
\centering
\includegraphics[width=4in,clip]{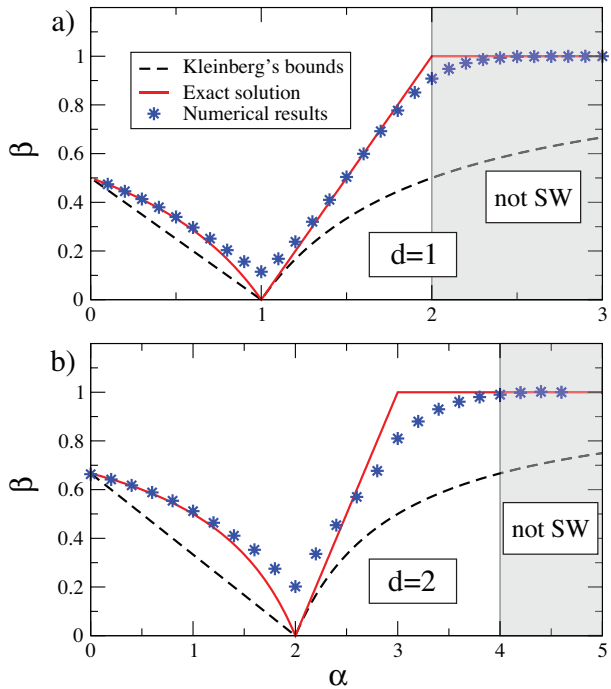} 
\caption{{\bf Asymptotic exponents of the greedy distance.} Comparison
  between the lower bounds found in the original reference (black
  dashed lines) and the exact solution presented here (red solid
  lines) in (a) $d=1$ and (b) $d=2$. The exact solutions are verified
  by numerical results (blue stars). The shadowed region ($\alpha > 2$
  in $d=1$ and $\alpha>4$ in $d=2$) is outside of the small-world
  regime.}
\label{Fig1}
\end{figure}

The simple decentralized algorithm considered in~\cite{Kleinberg2000}
is of greedy nature: starting from a given node, at every step the
algorithm chooses the edge with the end-node which is closer, in
euclidean or lattice distance, to the selected destination; such scheme guarantees
that the destination is always certainly reached. Using arguments from
probability theory, it was possible to find, in $d=2$, a lower bound
for the average number of steps $\tau(L)$ that are necessary to
connect nodes separated by a distance proportional to the lattice linear size 
$L$: $\tau(L)
\leq L^\beta$, with $\beta = (2-\alpha)/3$ if $0 \leq \alpha < 2$ and
$\beta = (\alpha-2)/(\alpha-1)$ if $\alpha > 2$ (see Fig.\ref{Fig1},
dashed lines). In the case $\alpha=2$ the upper bound $\tau(L) \leq
(\ln L)^2$ was found instead.  As a consequence, the best choice for
an optimized navigability of a SW network using a decentralized
algorithm would be $\alpha=2$ in $d=2$ and, more generally, $\alpha=d$
in $d$ dimensions~\cite{Kleinberg00,Kleinberg2000}.

In what follows, we derive the exact asymptotic behavior of $\tau(L)$
in any dimensions. We resort to the same implicit assumption
already used in the probabilistic approach
in~\cite{Kleinberg00}: we treat the algorithm as a stochastic Markov
process, by looking simultaneously at all possible network
realizations~\cite{Higham2006}.  If at a given stage the message is at
site $i$, at lattice distance $d_i$ from the target, at the next step
it will surely be at a site $j$, with distance $d_j < d_i$, due to the
greediness of the algorithm.  More in detail, if $j$ is a nearest
neighbor of $i$, then their connecting edge is chosen only if there
are no shortcuts from $i$ to sites $k$ with $d_k < d_j$. If, instead,
such a useful shortcut exists, the greedy algorithm chooses it over
the nearest neighbor connection.  In general, we can write the
following recursive relation for the average number of steps from a
site $i$ to the destination
\begin{equation}
\tau(i) = \sum_{j} p_{i \to j} (\tau(j)+1)
\label{tau equation - lattice}
\end{equation}
where the probability $p_{i \to j}$ depends on the presence and
greedy-usefulness of shortcuts. In equation (\ref{tau equation -
lattice}) we assume, following~\cite{Kleinberg2000},
that the time it takes to travel a network edge,
be it a shortcut or a link of the underlying lattice, is $1$.

Equation (\ref{tau equation - lattice}) can be easily solved numerically,
on a lattice, by recursion.  However, it is more instructive to take its
continuous space limit, where a lattice site $i$ is mapped onto a
position $\vec{r}$ and the lattice spacing vanishes. 
Equation (\ref{tau equation - lattice}) in $d$ dimensions then becomes 
(for a complete derivation see the auxiliary material~\cite{auxmat})
\begin{widetext}
\begin{equation}
\tau'(r) = 1 - q \mathcal{N} \tau(r) \int \mathrm{d}\Omega \int_0^{2 r \cos\theta} 
(\epsilon+l)^{-\alpha} l^{d-1} \mathrm{d}l + 
 q \mathcal{N} \int \mathrm{d}\Omega \int_0^{2r \cos\theta}
(\epsilon + l)^{-\alpha} l^{d-1} \tau\left(\sqrt{l^2+r^2-2rl\cos\theta}\right) \mathrm{d}l
\label{integro-differential equation}
\end{equation}
\end{widetext}
with $\epsilon$ a short-lengthscale cutoff that avoids divergences
when $l\to0$ ($\epsilon$ was clearly not necessary on a lattice).
In equation (\ref{integro-differential equation}), $\int \mathrm{d}\Omega$
is the integral over the $d$-dimensional hypersolid angle and it is
parametrized, among others, by the azimuthal angle $\theta$,
which covers only half of the hypersphere~\cite{auxmat}; $q$ is the linear probability density of shortcuts. Clearly
it takes no time to reach the destination starting from itself, and consequently
$\tau(0) =0$. Given the complete isotropy of the problem, $\tau(r)$ depends only
on $r=|\vec{r}|$.

By setting the scaling function form
\begin{equation}
\tau(r) = K^{-1} f(K r)
\end{equation} 
(the value of $K$ to be discussed
below case by case), it is possible to obtain, after some
manipulations~\cite{auxmat}, an integral equation for
the derivative of $f(x)$,
\begin{widetext}
\begin{equation}
f'(x) = 1-x^{d+1-\alpha} \int \mathrm{d}\Omega \int_0^{2\cos\theta} 
I(y,K\epsilon/x) \frac{y-\cos \theta}{\sqrt{y^2+1-2y\cos \theta}}f'\left(x \sqrt{y^2+1-2y\cos\theta}\right)
\mathrm{d}y
\label{integral equation}
\end{equation}
\end{widetext}
where $I(y,K\epsilon/x) = \int_0^{y}
\left(K\epsilon/x+z\right)^{-\alpha} z^{d-1} \mathrm{d}z$.  To obtain
the asymptotic behavior of $f(x)$ we rely on simple considerations.
Since $\tau(r)$ cannot decrease with the distance $r$ from the
target, $f(x)$ is non-decreasing and therefore $f'(x)$ is
non-negative. Moreover, $\mathcal{I} (y,K\epsilon/x)$ is positive and
thus from (\ref{integral equation}) we obtain $f'(x)\leq1$. It is then
possible to show that if $\alpha<d+1$, asymptotically $f'(x)$ depends
only on $\alpha$ and $f'(x) \sim x^{-(d+1-\alpha)}$. If instead
$\alpha \geq d+1$, $f'(x)$ is not universal anymore~\cite{auxmat} 
and $f'(x) \to 1/[1+c(\alpha,\epsilon,d)q]$, where $c(\alpha)$ is a
constant depending only on $d$ and $\alpha$.
 
\begin{figure}[h!] 
\centering
\includegraphics[width=3in,clip]{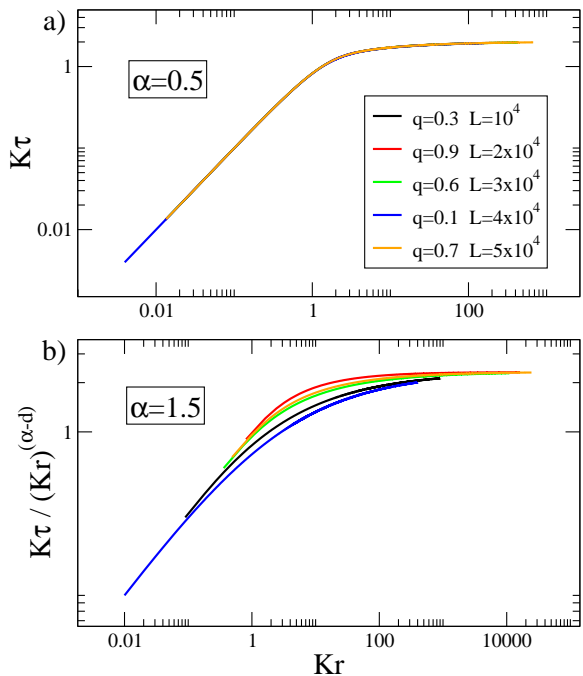} 
\caption{{\bf Universality of greedy distances.} The numerical results
  confirm the ansatz $\tau(r) = K^{-1} f(Kr)$ in $d=1$. The collapses
  obtained once $K \tau(r)$ is plotted as a function of $Kr$ for a
  fixed $\alpha$ but for different $(q,L)$ pairs show that
  asymptotically ($Kr \to \infty$) $f(Kr)$ is a universal function
  that only depends on $\alpha$.}
\label{Fig2}
\end{figure}

In the following we analyze the five cases $0 \leq \alpha<d$,
$\alpha=d$, $d<\alpha<d+1$, $\alpha=d+1$ and $\alpha >d+1$ separately:
\begin{itemize}
\item {\it $0 \leq \alpha < d$}: In this case the normalization is
  $\mathcal{N}=(d-\alpha)/L^{d-\alpha}$ and
  $K=\left[q(d-\alpha)\right]^{1/(d+1-\alpha)}
  L^{-(d-\alpha)/(d+1-\alpha)}$ to the leading order in $L$.  Since
  $f'(x) \sim 1/x^{(d+1-\alpha)}$, $f(x)$ converges asymptotically to
  a constant and consequently $\tau(L) \sim
  L^{(d-\alpha)/(d+1-\alpha)}/q^{1/(d+1-\alpha)}$ for large values of
  $L$.

\item {\it $\alpha = d$}: The normalization constant is $\mathcal{N} =
  \ln L$ to the leading order in $L$.  This implies $f'(x) \sim 1/x$,
  so that $\tau(r) \sim \left( \ln L \right)^2 /q$ for
  large values of $L$. This results coincides with the upper bound found in~\cite{Kleinberg2000}.

\item {\it $d \leq \alpha < d+1$}: The normalization does not depend, to the leading order,
on $L$. In this case $f'(x) \sim 1/x^{(d+1-\alpha)}$ is non-integrable and
therefore $f(x)$ diverges as $x^{\alpha-d}$, and asymptotically
$\tau(L) \sim L^{\alpha-d}/q$.

\item {\it $\alpha = d+1$}: Again the normalization does not depend on
  $L$.  After some calculations~\cite{auxmat} it is
  possible to show that $\tau(r) \sim L/(q\ln L)$.

\item {\it $\alpha > d+1$}: Once more the normalization does not
  depend on $L$. It is easy to show that $f'(x)\to 1/[1+c(\alpha,
    \epsilon, d)q]$ and thus $\tau(L) \sim L/(1+c(\alpha, \epsilon, d)q)$
  asymptotically.

\end{itemize}
The above results are summarized in Table \ref{Table}, and one major 
conclusion that can be drawn at this stage is that a greedy algorithm can significantly outperform the simple lattice distance only if the shortcut length distribution
has diverging first moments in the $L\to \infty$ limit.  The analytical
predictions, shown in Fig.\ref{Fig1} (red solid lines) for $d=1,2$,
are compatible with the lower bounds found in~\cite{Kleinberg2000}.
We have also numerically solved equation (\ref{tau equation -
lattice}) on a lattice, and using lattice distances, in $d=1,2$ and the estimates for the exponent $\beta$ agree
with the analytical predictions (Fig.\ref{Fig1}, blue stars), apart
from small discrepancies due to the not-yet fully achieved asymptotic
limit, an effect that is expectedly more important in higher
dimensions. Furthermore, in $d=1$ we have verified that, as long as
$\alpha<2$, the $\tau(r)$ curves for different values of $q$ and of
$L$, but for the same value of $\alpha$, do collapse onto each other
asymptotically once $K \tau(r)$ is plotted as a function of $Kr$ with
the appropriate values of $K$ (see Fig.\ref{Fig2}). This result
confirms the asymptotic universality of $f'(x)$ when $\alpha< d+1$.

The results described above are necessary but not yet sufficient to
thoroughly address the navigability of SW networks. Indeed, changing
the value of $\alpha$ while keeping the value of $q$ fixed, as
in~\cite{Kleinberg2000}, is only one of the possible ways to compare
the navigability of SW networks. As a matter of fact, it has been
shown that, as long as $0\leq \alpha <2d$, it is possible to make a
$d$-dimensional lattice {\it small} by letting the shortcut
probability $q$ depend on $\alpha$~\cite{Petermann2006}. In
particular, the shortcut probability $q$ marking the crossover from
the euclidean to the small world regime has the form $q \sim L^{-d}$
if $0 \leq \alpha < d$, $q\sim \ln L L^{-d}$ if $\alpha=d$ and $q\sim
L^{\alpha-2d}$ if $d<\alpha<2d$. Thus, keeping $q$ fixed for different
values of $\alpha$ would pit against each other networks that are
intrinsically differently {\it small}.

To alleviate such bias, a more appropriate comparison should take into
account how $q$ must change with $\alpha$.  In order to do so, we can
start from a given value of the shortcut probability $q(d)$ at
$\alpha=d$, where purportedly navigation is easier.  Then, as $\alpha$
moves away from $d$, we let this value change according to $q(\alpha)
= q(d)/\ln L$ if $\alpha < d$ and $q(\alpha) = q(d) L^{\alpha-d}/\ln
L$ if $d < \alpha < 2d$.  Since our derivation of the asymptotic
behavior of $\tau(l)$ also provides the precise form of the $q$-dependent
prefactors, we can obtain the correct expected asymptotics of
$\tau(L)$ when $q$ is allowed to appropriately change: $\tau(L) \sim
\left(L^{d-\alpha} \ln L \right)^{1/(d+1-\alpha)}$ if $0 \leq \alpha <
d$, $\tau(L)\sim (\ln L)^2$ if $\alpha=d$, $\tau(L) \sim \ln L$ if $d
< \alpha < d+1$, $\tau(L) \sim \mathrm{const}$ if $\alpha=d+1$ and
$\tau(L) \sim L^{-[\alpha-(d+1)]}$ if $d+1 < \alpha < 2d$. The last
result is apparently paradoxical, because it predicts the navigation
time to become smaller for larger systems. Actually, using the results
obtained in~\cite{Petermann2006}, it is possible to show that if
$q(\alpha) = q(d) L^{\alpha-d}/\ln L$, the average path length of a
small-world actually decreases with $L$ for fixed $\alpha$, even
faster than using a greedy algorithm~\cite{auxmat}. 
It is thus not surprising that also within the greedy
framework larger values of $\alpha$, with the corresponding increase
of $q$, lead to shorter navigation times. Considering all the
results summerized in Table \ref{Table}, we conclude that if
we take the necessary change of $q$ into account any $d < \alpha < 2d$
outperforms the case $\alpha=d$.

To gain a definitive insight into the properties of such class of SW
networks, we finally consider the total shortcut length per node
$\mathcal{L}=q(\alpha,L) \bar{l}(\alpha,L)$ ($\bar{l}(\alpha,L)$ being
the average shortcut length), for different values of
$\alpha$. $\mathcal{L}$ can be associated to the amount of resources
(cost) needed to set up such networks. We have $\mathcal{L} =
q(d)L/\ln L$ for $0<\alpha < d+1$ and $\mathcal{L} = q(d) \bar{l}
L^{\alpha-d}/\ln L$ for $\alpha \geq d+1$. We observe that the value
of $\mathcal{L}$ is the same for all $0 \leq \alpha < d+1$, and lower
than that for $d+1 \leq \alpha < 2d$.

The results presented above reveal a new picture
for the navigability of SW networks, which is richer and more variegate
than previously outlined: navigability depends on several parameters,
and the optimal choice depends on the criteria that have to be
satisfied. If the average number $q$ of shortcuts per node 
must be kept constant, at the cost
of making networks characterized by different values of $\alpha$ 
differently {\it small}, then our exact results confirm, while making more precise,
the findings of~\cite{Kleinberg2000}. If on the other hand a fairer comparison 
of small-world networks is desired, then the necessary change of $q$ as a function of $\alpha$ must be taken into account and any $\alpha > d$ outperforms $\alpha = d$,
with optimal navigability reached for $\alpha=2d$. If finally the amount of resources
needed to set up the networks becomes a crucial factor, the whole range $\alpha \in
(d, d+1)$ is where navigability is optimal and cheapest.

Outlining the dependence of navigability on various criteria, as highlighted in this work,
is important in order to understand the architecture of networks where the small-world
feature is believed to be crucial. Indeed, it can be expected that virtual connections,
such as hyperlinks between web-pages, will bear almost no signature of cost-effects,
but might be limited in number to a few units; contrary-wise, the number of connections per neuron in the brain can be extremely high, but establishing and maintaining them is surely resource expensive. Thus different systems might have achieved different layouts to enhance their navigability while respecting given sets of constraints.
Keeping in mind all these criteria can be of paramount relevance given the growing interest in the greedy navigability of real networks~\cite{Boguna2009}.

The authors would like to thank the Swiss National Research Foundation (200020Ð116286) for financial support.


\begin{table}[h!]
\begin{tabular}{c c | c  | c c }
& & $\tau(L,q)$ & & $\tau(L,q(L,\alpha))$  \\ \hline
 $0 \leq \alpha < d$ & & $L^{\frac{d-\alpha}{d+1-\alpha}}$ & & 
 $\left(\frac{L^{d-\alpha}}{\ln L}\right)^{\frac{1}{d+1-\alpha}}$    \\ \hline
 $\alpha = d$ & & $(\ln L)^2 $ & & $(\ln L)^2$  \\ \hline
 $d < \alpha < d+1$ & & $L^{\alpha-d}$  & & $\ln L$ \\ \hline    
 $\alpha = d+1$ & & $L/\ln L$  & & const \\ \hline    
 $ d+1 < \alpha < 2d$ & & $L$ & & $L^{-[\alpha-(d+1)]}$ \\ \hline
\end{tabular}
\caption{{\bf Asymptotic navigation times as a function of the
exponent $\alpha$.} The column $\tau(L,q)$ reports the navigation time
at constant shortcut density, whereas the column $\tau(L,q(L,\alpha))$
reports the navigation time when the shortcut density is allowed to
change so to keep the system in the {\it small-world} regime.}
\label{Table}
\end{table}

\widetext
\newpage

\begin{center}
\huge{\bf Supplementary Material}
\end{center}

\section{Stochastic process formulation}

Kleinberg's problem can be formulated as a stochastic process when
analyzed on the whole small-world network ensamble.  We consider a
walker that starts from site $i$ at distance $d_i$ from the
destination.  At the next step it will move differently on every
network realization in the ensemble. Thus it will be on a site $j$ at
distance $d_j$ from the destination with a probability $q$, that is
the probability that site $i$ has a shortcut, times the probability
that there is a shortcut from $i$ to $j$, times the probability such
shortcut can be used by the greedy algorithm, {\it i.e.} that $d_j <
d_i$. When $j$ is a nearest neighbor of $i$ (with $d_j = d_i-1$), the
probability of the walker to be in $j$ is given by the probability
that $i$ does not have any shortcuts or that it does but that they end
at distances from the destination larger than $d_j$. In that case the
algorithm chooses an edge of the underlying lattice. It is then
possible to formulate the evolution of the probability distribution of
the walker's position as
\begin{equation}
P_k(t+1) = \sum_j p_{j \to k} P_j(t)
\end{equation}
with initial conditions $P_k(0) = \delta_{i,k}$ if the walker starts
from $i$.  Given the nature of the greedy algorithm, the destination
will be reached with certainity in a number of steps that is at most
$d_i$.

Once the stochastic formulation is established, it is then easy to
derive that the average trip time $\tau_i$ from site $i$ to the
destination can be expressed recursively using the average trip time
from every other node closer to the destination than $i$:
\begin{equation}
\tau_i = \sum_j p_{i \to j} \left(\tau_j+1\right)
\label{eq lattice}
\end{equation}
where the $1$ in the parenthesis indicate the time it takes to go from
site $i$ to site $j$ if there is a usable edge (shortcut or otherwise)
from $i$ to $j$, which, as explained above, occurs with probability
$p_{i \to j}$.  Equation (\ref{eq lattice}) can be easily solved with
the aid of the computer, but it is extremely demanding on time and
memory resources for lattices with dimension $d \geq 2$ if the
asymptotic limit has to be reached.

\section{Continuous space formulation of equation (\ref{eq lattice})}

The continuous space formulation of equation (\ref{eq lattice}) is
easily obtained as follows. For reasons that will be clear in the
following, shortcuts do not connect lattice sites, but rather the
shortcut associated to a site go from the lattice edges connecting
it to its neighbors to a lattice node. A small-world network obtained
in this way is clearly as small-world as a network obtained with
shortcuts from site to site.

Without affecting the generality of the results, from now on we shall
assume that the destination is the origin. We also assume spatial
isotropy of the process. Say that the walker starts from a point
$\vec{r}$, at distance $r$ from the destination. Since it obeys a
greedy scheme, it will always try to approach the destination as much
as possible. Thus, in the absence of any useful shortcut, it will move
along the ray defined by $\vec{r}$. Let's examine what may happen
while it moves along the ray for an infinitesimal distance
$\mathrm{d}r$. The point at distance $r-\mathrm{d}r$ will be reached
only if 1) no shortcuts are encountered, with a probability $1-qdr$
(where $q$ is now the linear density of shortcut probability, which
also explains why the starting points of shortcuts are more
appropriately associated to lattice edges in this context), or 2) if
shortcuts are encountered (probability $q\mathrm{d}r$) but none of
them are useful. The expected time to reach the
destination, in $d$ dimensions, can then be written as
\begin{equation}
\tau(r) = \left[ \left(1-q\mathrm{d}r\right) + q \mathrm{d}r
\left(1-\int \mathrm{d}\Omega \int_0^{2r \cos\theta} \mathcal{N}
(\epsilon+l)^{-\alpha} l^{d-1} \mathrm{d}l \right) \right]
\left(\tau(r-\mathrm{d}r) + \frac{\mathrm{d}r}{v}\right) + \dots
\label{first part}
\end{equation}
where the dots indicate the contribution of {\it useful} shortcuts
encountered along $\mathrm{d}r$, that we will discuss
below. $\mathcal{N} (\epsilon+l)^{-\alpha}$ is the shortcut length
probability distribution and $\epsilon$ is a short-distance cutoff
introduces in order to avoid divergences when $l\to0$ if $\alpha \geq
d$.
\begin{figure}[t!] 
\centering
\includegraphics[width=5in,clip]{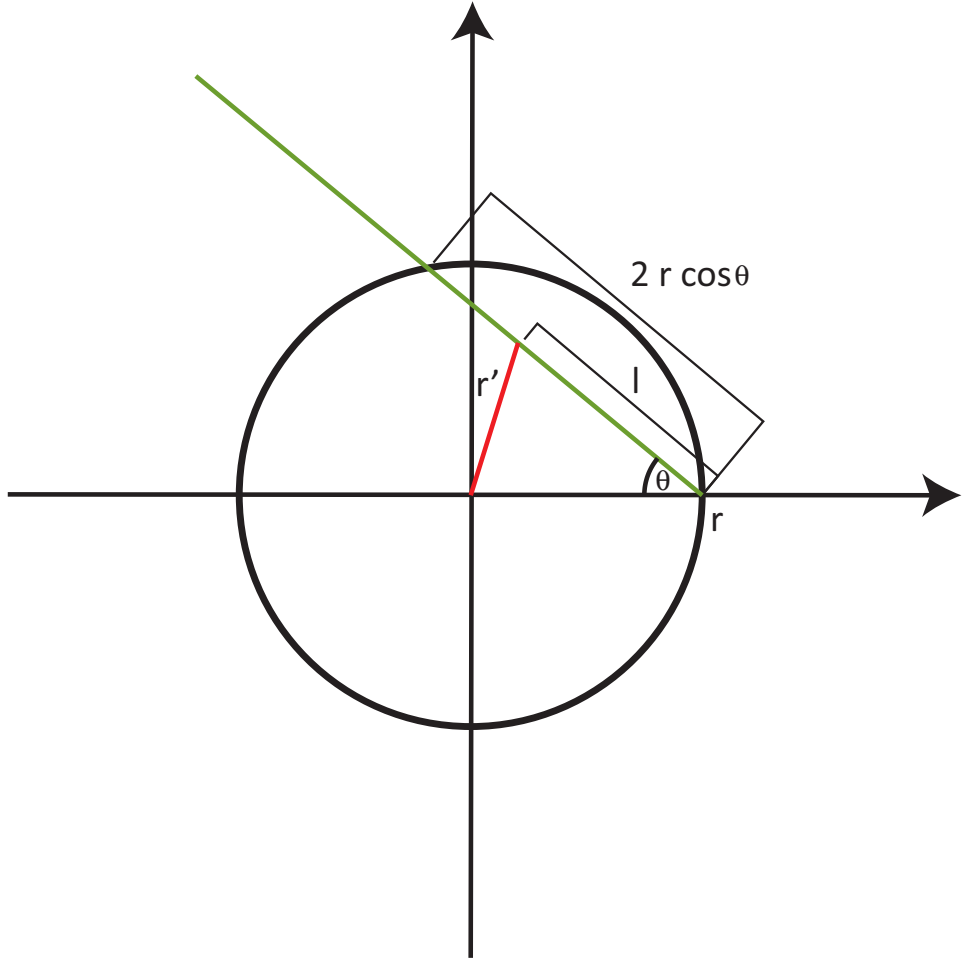} 
\caption{Graphical representation of the relations between $\vec{r}$,
$\vec{r'}$ and $l$.}
\label{SuppMatFig1}
\end{figure}

The double integral in equation (\ref{first part}) is the probability
for a shortcut encountered between $r$ and $r-\mathrm{d}r$ to fall
inside a sphere of radius $r$, thus closer to the destination than
$r$.  It is composed of an angular integral over all the angles
describing a (hyper)spherical parametrization of the space centered on
$\vec{r}$ (with $\theta \in [0,\pi/2]$ in the integral over
$\mathrm{d}\Omega$), and of a radial part $l$ centered on
$\vec{r}$. The triangle defined by $\vec{r}$, $\vec{l}$ and
$\vec{r}-\vec{l}$ identifies a two-dimensional plane in any dimension.
Given the isotropy of the space, there is full rotational invariance
around the ray $\vec{r}$, hence $\theta$ is the only angular variables
of the integral over $\mathrm{d}\Omega$ that affects the radial
integral.  The reason for this peculiar choice of the parametrization
will be explained below.

The term $\tau(r-\mathrm{d}r) + \frac{dr}{v}$ is the expected time to
reach the destination from a distance $r-\mathrm{d}r$ augmented of the
time to reach $r-\mathrm{d}r$ from $r$, which we assume to be
proportional to $\mathrm{d}r$ and to depend on an intrinsic velocity $v$.
In what follows, we fix $v=1$.

Equation (\ref{first part}) is not complete yet. The term representing the
contribution from useful shortcuts is given by:
\begin{equation}
\tau(r) = \dots + q \mathrm{d}r \int \mathrm{d}\Omega \int_0^{2r
\cos\theta} \mathcal{N} (\epsilon+l)^{-\alpha} l^{d-1}
\left(\tau\left(\sqrt{l^2+r^2-2lr\cos\theta}\right) +
\frac{\mathrm{d}r}{v}\right) \mathrm{d}l
\label{second part}
\end{equation}
In complete analogy with our assumption, in equation (\ref{eq
  lattice}), that it takes exactly the same unitary time to travel
along a lattice edge or a shortcut, the time to reach $r-\mathrm{d}r$
from $r$ is the same in (\ref{first part}) and (\ref{second part})
(with $v=1$ in the latter).

By dividing both terms by $\mathrm{d}r$ and taking the limit
$\mathrm{d}r \to 0$, we obtain the continuos space limit presented in
the main text:
\begin{equation}
\tau'(r) = 1 - q \mathcal{N} \tau(r) \int \mathrm{d}\Omega \int_0^{2r
\cos\theta} (\epsilon+l)^{-\alpha} l^{d-1} \mathrm{d}l + q
\mathcal{N}\int \mathrm{d}\Omega \int_0^{2r \cos\theta}
(\epsilon+l)^{-\alpha} l^{d-1}
\tau\left(\sqrt{l^2+r^2-2lr\cos\theta}\right) \mathrm{d}l \quad.
\label{integro-differential equation}
\end{equation}

\section{Rescaling of equation (\ref{integro-differential equation})}

When we introduce the assumption $\tau(r) = K^{-1} f(Kr)$ the left hand side
of equation (\ref{integro-differential equation}) becomes $\tau'(r) =
f'(Kr)$. After some manipulations, we can write:
\begin{eqnarray}
f'(Kr) &=& 1 - q \mathcal{N} K^{-(d+1-\alpha)} f(Kr) \int
\mathrm{d}\Omega \int_0^{2r \cos\theta} (K\epsilon+Kl)^{-\alpha}
(Kl)^{d-1} \mathrm{d}(Kl) + \nonumber \\ &+& q \mathcal{N}
K^{-(d+1-\alpha)} \int \mathrm{d}\Omega \int_0^{2r \cos\theta}
(K\epsilon+Kl)^{-\alpha} f\left(\sqrt{K^2 l^2+K^2 r^2-2 (Kl) (Kr)
\cos\theta}\right) (Kl)^{d-1} \mathrm{d}(Kl) \quad .
\label{integro-differential equation 2}
\end{eqnarray}

A simple change of variable $y=Kl/Kr$ in the integral leads to
\begin{eqnarray}
f'(Kr) &=& 1 - q\mathcal{N} K^{-(d+1-\alpha)} (Kr)^{d-\alpha} f(Kr)
\int \mathrm{d}\Omega \int_0^{2\cos\theta}
\left(\frac{K\epsilon}{Kr}+y \right)^{-\alpha} y^{d-1} \mathrm{d}y +
\nonumber \\ &+& q \mathcal{N} K^{-(d+1-\alpha)} (Kr)^{d-\alpha} \int
\mathrm{d}\Omega \int_0^{2\cos\theta} \left(
\frac{K\epsilon}{Kr}+y\right)^{-\alpha}
f\left(Kr\sqrt{y^2+1-2y\cos\theta}\right) y^{d-1} \mathrm{d}y \quad .
\label{integro-differential equation 3}
\end{eqnarray}
Now we can set the value of $K$ according to $q \mathcal{N}
K^{-(d+1-\alpha)}=1$. Once the corresponding expression of $K$ is
introduced, equation (\ref{integro-differential equation 3}) depends
only on $Kr$. We can, then, change the general variable to
$x=Kr$ and obtain the equation
\begin{eqnarray}
f'(x) &=& 1 - x^{d-\alpha} f(x) \int \mathrm{d}\Omega
\int_0^{2\cos\theta} \left(\frac{K\epsilon}{x}+y \right)^{-\alpha}
y^{d-1} \mathrm{d}y + \nonumber \\ &+& x^{d-\alpha} \int
\mathrm{d}\Omega \int_0^{2\cos\theta} \left(\frac{K\epsilon}{x}+y
\right)^{-\alpha} f\left(x\sqrt{y^2+1-2y\cos\theta}\right) y^{d-1}
\mathrm{d}y \quad .
\label{integro-differential equation 4}
\end{eqnarray}

\section{Transforming the integro-differential equation into an integral equation}

By calling $I(y,K \epsilon/x)$ the primitive of
$\left(\frac{K\epsilon}{x}+y \right)^{-\alpha} y^{d-1}$, equation
(\ref{integro-differential equation 4}) becomes
\begin{eqnarray}
f'(x) &=& 1 - x^{d-\alpha} f(x) \int \mathrm{d}\Omega \left[
I(2\cos\theta,K\epsilon/x)-I(0,K\epsilon/x) \right] + \nonumber \\ &+&
x^{d-\alpha} \int \mathrm{d}\Omega \int_0^{2\cos\theta}
\left(\frac{K\epsilon}{x}+y \right)^{-\alpha} y^{d-1}
f\left(x\sqrt{y^2+1-2y\cos\theta}\right) \mathrm{d}y \quad .
\label{integro-differential equation 5}
\end{eqnarray}

Integration by parts the last integral in equation
(\ref{integro-differential equation 5}) gives
\begin{eqnarray}
f'(x) &=& 1 - x^{d-\alpha} f(x) \int \mathrm{d}\Omega \left[
I(2\cos\theta,K\epsilon/x)-I(0,K\epsilon/x) \right] + \nonumber \\ &+&
x^{d-\alpha} \int \mathrm{d}\Omega
\left[I(2\cos\theta,K\epsilon/x)-I(0,K\epsilon/x)\right]f(x) +
\nonumber \\ &-& x^{d+1-\alpha}\int \mathrm{d}\Omega
\int_0^{2\cos\theta} I(y, K
\epsilon/x)\frac{y-\cos\theta}{\sqrt{y^2+1-2y\cos\theta}}
f'\left(x\sqrt{y^2+1-2y\cos\theta}\right) \mathrm{d}y \quad
\label{integro-differential equation 6}
\end{eqnarray}
which simplifies to
\begin{eqnarray}
f'(x) &=& 1 - x^{d+1-\alpha}\int \mathrm{d}\Omega \int_0^{2\cos\theta}
I(y,K \epsilon/x) \frac{y-\cos\theta}{\sqrt{y^2+1-2y\cos\theta}}
f'\left(x\sqrt{y^2+1-2y\cos\theta}\right)\mathrm{d}y \quad .
\label{integral equation 1}
\end{eqnarray}
This is an integral equation for $f'(x)$.  The properties of the
kernel $I(y,K\epsilon/x)$ are simple: it is always positive because it
is the integral of a positive function, and $I(0,K\epsilon/x) \sim
(K\epsilon/x)^{d-\alpha}$.  Thus, in the asymptotic limit $x\to
\infty$, $I(y,K\epsilon/x)$ develops an integrable singularity for
$\alpha<d+1$ and it converges to a well-behaved kernel. In what
follows we can safely disregard the short-distance cutoff.  For
$\alpha \geq d+1$, the singularity is non-integrable and it will
require a special treatment.

Equation (\ref{integral equation 1}) shows that, since
we can safely take the $K \epsilon/x \to 0$ limit when $\alpha < d+1$,
$f'(x)$ must obey an asymptotically universal form that depends only
on $\alpha$. We have numerically verified that this is indeed the case
(see main text). In the case $\alpha \geq d+1$, instead, the $K\epsilon/x$ term
dominates and $f'(x)$ will not be universal any more.

\section{Asymptotic behavior of $f'(x)$}

We want to determine the asymptotic behavior of $f'(x)$.  Clearly
$\tau(r)$ must be a non decreasing function of $r$, since it obviously takes
longer to reach the destination starting from increasingly
farther distances.  Thus, $f'(x)$ is non negative. Consequently, equation
(\ref{integral equation 1}) tells us that $f'(x) \leq 1$. We can, then, rewrite
equation (\ref{integral equation 1}) as
\begin{equation}
f'(x) = 1- \int \mathrm{d}\Omega \int_0^{2\cos\theta} I(y,K
\epsilon/x)\frac{y-\cos\theta}{\left(\sqrt{y^2+1-2y\cos\theta}\right)^{d+2-\alpha}}
\left[ \left(x\sqrt{y^2+1-2y\cos\theta}\right)^{d+1-\alpha}
f'\left(x\sqrt{y^2+1-2y\cos\theta}\right) \right] \mathrm{d}y \quad
\label{integral equation 2}
\end{equation}

\subsection{The case $\alpha < d+1$}

If $\alpha<d+1$ it is easy to show that the only ansatz for the 
behavior of $x^{d+1-\alpha} f'(x)$ that
does not lead to contradicting conclusions is that 
\begin{equation}
x^{d+1-\alpha} f'(x) \xrightarrow{x \to \infty}
\mathcal{C}=\left\lbrace\int \mathrm{d}\Omega \int_0^{2\cos \theta}
I(y,K \epsilon/x)\frac{y-\cos\theta}{\left(\sqrt{y^2+1-2y
\cos\theta}\right)^{d+2-\alpha}} \mathrm{d}y\right\rbrace^{-1} \quad
\label{integral equation 3}
\end{equation}
so that $f'(x) \sim \mathcal{C}/x^{d+1-\alpha}$.  Indeed, should
$f'(x)$ decrease faster than that, then the whole integral would
become increasingly negligible and $f'(x)$ would be asymptotically
equal to $1$, against its fast decay assumption. Should
$f'(x)$ decay slower than $\mathcal{C}/x^{d+1-\alpha}$, then the
integral would asymptotically diverge and $f'(x)$ would become
negative, implying a non-physical decrease of $\tau(r)$ for larger
distances.

\subsection{The case $\alpha = d+1$}

In this case $K$ is not set by any equation, hence we can choose
$K=1$. This implies that in the rescaling of the continuos limit we
can simply set $\tau(r) = f(x)$ and $r=x$. The integral equation
becomes (already in the limit when $x\to \infty$)
\begin{eqnarray}
f'(x) &=& 1 - q\mathcal{N} \int \mathrm{d}\Omega \int_0^{2\cos\theta}
\delta(y) \frac{y-\cos\theta}{\sqrt{y^2+1-2y\cos\theta}}
f'\left(x\sqrt{y^2+1-2y\cos\theta}\right)\mathrm{d}y \quad .
\label{integral equation spec}
\end{eqnarray}
Since asymptotically $\int_0^{2\cos\theta} I(y,\epsilon/x)\mathrm{d}y
\sim \ln{x/\epsilon}$, we have $f'(x) = 1/[1+q c(\alpha,d) \ln
x]$. Thus $\tau(r) \sim \frac{1}{q\mathcal{N}c(\alpha,d)}\frac{x}{\ln
x}$.

\subsection{The case $\alpha>d+1$}

If $\alpha \geq d+1$, the kernel diverges as
$(x/K\epsilon)^{\alpha-d}$ for $x\to \infty$.  We thus divide and
multiply it for its integral between $0$ and $2\cos\theta$ to obtain
\begin{eqnarray}
f'(x) &=& 1 - x^{d+1-\alpha} \left[\int_0^{2\cos\theta}
I(y,K\epsilon/x)\mathrm{d}y\right] \int \mathrm{d}\Omega
\int_0^{2\cos\theta} \mathcal{I}(y,K
\epsilon/x)\frac{y-\cos\theta}{\sqrt{y^2+1-2y\cos\theta}}
f'\left(x\sqrt{y^2+1-2y\cos\theta}\right)\mathrm{d}y \nonumber \\ &&
\label{integral equation 4}
\end{eqnarray}
where $\mathcal{I}(y,K\epsilon/x)= I(y,K\epsilon/x)/
\left[\int_0^{2\cos\theta}
I(y,K\epsilon/x)\mathrm{d}y\right]$. Interestingly,
$\mathcal{I}(y,K\epsilon/x) \to \delta(y)$ as $x \to \infty$ and thus
equation (\ref{integral equation 4}) reduces to
\begin{equation}
f'(x) = 1- x^{d+1-\alpha} \left[\int_0^{2\cos\theta}
I(y,K\epsilon/x)\mathrm{d}y\right] f'(x) \quad.
\label{equation for alpha >= d+1}
\end{equation}
Since $\int_0^{2\cos\theta} I(y,K\epsilon/x)\mathrm{d}y \sim
(x/K\epsilon)^{\alpha-d-1}$, from equation (\ref{equation
for alpha >= d+1}) we obtain that $f'(x) \sim 1/(1+c(\alpha,\epsilon,d) q)$,
where $c(\alpha, \epsilon, d)$ is a constant that depends only on
$\alpha$, $\epsilon$ and $d$. Thus, $f(x) \sim
x/(1+c(\alpha,\epsilon,d) q)$ and $\tau(r) \sim
r/(1+c(\alpha,\epsilon,d) q)$.

This result can be derived on a lattice by simple arguments. When
$\alpha > d+1$, the average shortcut length $\bar{l}$ is finite, so
that shortcuts simply redefine the elementary length-scale of the
system. The typical step length is, then, $l_{typ} = 1\cdot
\lbrace(1-q) + q[1-\tilde{c}(\alpha,\epsilon,d)]\rbrace + \bar{l}\cdot
q\tilde{c}(\alpha,\epsilon,d) =
1+(\bar{l}-1)q\tilde{c}(\alpha,\epsilon,d)$, where
$\tilde{c}(\alpha,\epsilon, d)$ is the probability that, if a shortcut
is present, it can be used by the greedy algorithm. $\bar{l} > 1$
because on a lattice there are no shortcuts of unitary length. The
average time to reach the destination, given that it increases
linearly, will be $r/l_{typ}$.

\section{Greedy and global path length in the case $\alpha > d+1$ and $q(\alpha) = q(d) L^{\alpha-d}/\ln L$ }

As outlined and numerically verified in [Petermann and De Los Rios,
Phys. Rev. E {\bf 73}, 026114 (2006)], the average path length $<L>$
for small world networks of linear size $L$ built according to the
present recipe (given $\alpha$ and $q$) obeys the scaling relation
\begin{equation}
<L> = L^* \mathcal{F}_\alpha\left(\frac{L}{L^*}\right)
\label{scaling from Petermann}
\end{equation}
where $L^* = q^{-1/d}$ if ($\alpha < d$) and $L^* =
q^{-1/(2d-\alpha)}$ if ($d <\alpha < 2d$).  The function
$\mathcal{F}_\alpha(x)$ is
\begin{equation}
\mathcal{F}_\alpha (x) = \left\{ 
\begin{array}{l l}
  x & \quad \mbox{if $x \ll 1$}\\
  (\ln x)^\gamma & \quad \mbox{if $x \gg 1$ }\\
\end{array} \right. 
\end{equation}
where the $x >> 1$ asymptotic behavior is a power of a logarithm (see
the above mentioned reference and also [Kosmidis, Havlin and Bunde,
Europhys. Lett. {\bf 82}, 48005 (2008)]).

Using $q=q(d) L^{\alpha-d}/\ln L$ (for $\alpha>d$) in equation 
(\ref{scaling from Petermann}) it is possible to obtain
\begin{equation}
<L> = L^{-\frac{\alpha-d}{2d-\alpha}}(\ln L)^{\frac{1}{2d-\alpha}} \mathcal{F}_\alpha 
\left(\frac{L^{\frac{d}{2d-\alpha}}} {(\ln L)^{\frac{1}{2d-\alpha}}}\right) \quad.
\label{average path length}
\end{equation}

Thus, since $\mathcal{F}_\alpha$ increases at most logarithmically,
the average path length decreases asymptotically as a power of $L$ as
$\alpha$ increases. The logarithmic decrease of the greedy path length
for $\alpha > d+1$ is thus compatible with this behavior and,
expectedly, the greedy path length remains larger than the global one.

\end{document}